# Simple model on collisionless thin-shell instability growth

**D. Doria[1], A. Bret[2,3], M. E. Dieckmann[4]**

[1]Centre for Plasma Physics, School of Mathematics and Physics, Queen's University of Belfast, Belfast BT7 1NN, United Kingdom

[2]ETSI Industriales, Universidad Castilla La Mancha, E-13 071 Ciudad Real, Spain

[3]Instituto de Investigaciones Energéticas y Aplicaciones Industriales, Campus Universitario de Ciudad Real, 13071 Ciudad Real, Spain

[4]Department of Science and Technology, Linkoping University, SE-60174 Norrkoping, Sweden

**(Dated: Nov 9, 2016)**

Thin-shell instability (TSI) is a process normally observed in astrophysical plasma on collisional scales. It is generated by an imbalance between the ram pressure of the inflowing upstream plasma and the downstream thermal pressure at the shock interface [1-4]. Recently, it has been shown by means of a particle-in-cell simulation that an analog process can destabilize a thin shell formed by two interpenetrating, unmagnetized, and collisionless plasma clouds [5]. A detailed mathematical description of the TSI would be rather involved and beyond the purpose of this manuscript. Here we rather intend to present a simple TSI model by only imposing the fulfillment of conservation laws. We shall only apply the laws of mass and momentum conservation and we will not take into account energy repartition inside the thin shell.

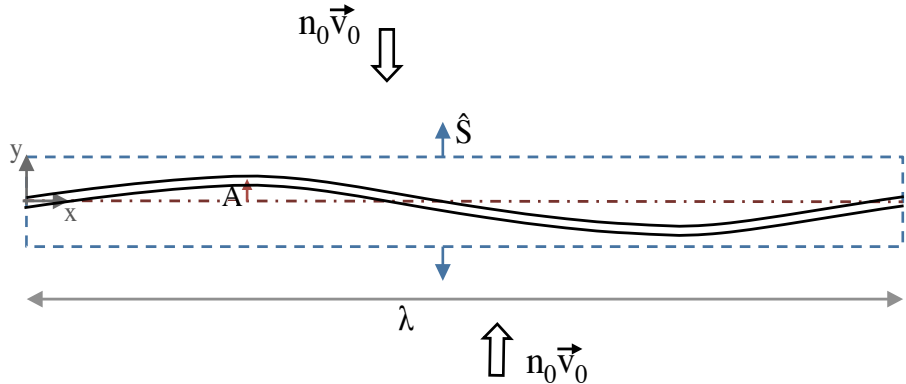

**Fig. 1.** Thin Shell Instability representation. The surface contour S initially coincides with the boundaries of the thin shell. The same condition is assumed to hold approximately at later time, under the small amplitude approximation $A \sim 0$. The total flux entering the thin shell is the sum of the fluxes $\mathbf{n_0 v_0}$ entering from both top and bottom sides of the shell.

The equations of continuity of mass and linear momentum in integral form are:

Mass conservation: $\qquad \frac{\partial}{\partial t} \iiint_V \rho \, dV = - \oiint_S \rho \vec{v} \cdot d\vec{S}$ (1)

Momentum Conservation: $\qquad \frac{\partial}{\partial t} \iiint_V \rho \vec{v} \, dV = - \oiint_S \rho \vec{v}\vec{v} \cdot d\vec{S}$, (2)

where $V$ is the shell volume and $S$ its surface contour (which has an indefinite extend $d$ along the z-axis), $\rho=mn$ is the mass density (with $m$ being the ion mass and $n$ the ion density) and $\vec{v}$ is the velocity of the element of mass $\rho dV$. The volume $V$ is filled at the rate $n_0\vec{v}_0$ from both the top and bottom surfaces, where $n_0$ and $\vec{v}_0$ are the background ion density and velocity in the shock frame respectively (Fig. 1).

Assuming the flux going outward $S$ to be negligible, (1) and (2) can be rewritten respectively as:

$$\frac{\partial N}{\partial t} = -\frac{1}{d}\oiint_S n\vec{v}\cdot d\vec{S} = 2n_0 v_0 \lambda \tag{3}$$

$$\frac{\partial \vec{P}}{\partial t} = -\frac{1}{d}\oiint_S mn\vec{v}\vec{v}\cdot d\vec{S} = 0 \;, \tag{4}$$

where $N$ and $\vec{P}$ are the total number of ions and linear momentum per unit length in the volume $V$ respectively. The Eq.s above show that the momentum is conserved, as can be deduced from the symmetry of the problem, and that the total ion number per unit length $N$ is varying linearly with time:

$$\Delta N = 2n_0 v_0 \lambda (t - t_0) \tag{5}$$

Assuming that the thin shell amplitude is initially growing as a sinusoidal perturbation $y = f[x,t]$ of wavelength $\lambda$, we can calculate the total length L of the shell as:

$$L = \int_0^\lambda \sqrt{1 + (\partial f[x,t]/\partial x)^2}\, dx \tag{6}$$

where:

$$f[x,t] = A[t]\sin\left[\frac{2\pi}{\lambda}x\right] \tag{7}$$

The length L results

$$L = \frac{2}{\pi}\sqrt{\lambda^2 + 4\pi^2 A[t]^2}\; E\left[1 - \frac{\lambda^2}{\lambda^2 + 4\pi^2 A[t]^2}\right] \tag{8}$$

where $E[m] = E\left[\frac{\pi}{2}|m\right]$ is the complete elliptic integral of the second kind. The shell is assumed to be thin, with mean thickness $\delta$ (i.e. $\delta(t) \ll \lambda \leq L$ at the formation stage) and ion density $\eta(t)$, therefore the to of the shell can be expressed as:

$$\eta L \delta \approx N = 2n_0 v_0 \lambda (t - t_0) + N_0 \tag{9}$$

where $N_0$ is the initial total ion number per unit length of the shell at time $t_0$. In order to find an explicit expression for $A[t]$ we expand (8) in series about the point $A[t_0]=0$:

$$L[A[t]] = \lambda + \frac{\pi^2 A[t]^2}{\lambda} + O[A[t]]^3 \tag{10}$$

Hence (9) can be rewritten as:

$$\eta \left(\lambda + \frac{\pi^2 A[t]^2}{\lambda}\right) \delta = 2 n_0 v_0 \lambda (t - t_0) + N_0 \tag{11}$$

At the time $t_0$, i.e. at the beginning of the perturbation, (11) gives:

$$\eta_0 \lambda \delta_0 = N_0 \tag{12}$$

where $\delta_0$ and $\eta_0$ are respectively the initial mean thickness and ion density of the shell before the perturbation begins.
Using (11) and (12), the amplitude $A$ can be rewritten as:

$$A[t] = \frac{\lambda}{\pi} \sqrt{\frac{2 n_0 v_0 \cdot (t - t_0) + (\eta_0 \delta_0 - \eta \delta)}{\eta \delta}} \tag{13}$$

where the $\delta$ and $\eta$ are generic functions of time.
If we evaluate the shell amplitude evolution at the formation stage and assume that the shell thickness and density are slowly varying with time, we can expand both in a series about $t_0$, and taking only the terms up to the first order we obtain:

$$\delta \approx \delta_0 + \delta_1 \cdot (t - t_0) \tag{14}$$
$$\eta \approx \eta_0 + \eta_1 \cdot (t - t_0), \tag{15}$$

And replacing these two last Eq.s into Eq.13 we obtain:

$$A[t] \approx \frac{\lambda}{\pi} \sqrt{\frac{\left(2 \frac{n_0 v_0}{\eta_0 \delta_0} - \left(\frac{\delta_1}{\delta_0} + \frac{\eta_1}{\eta_0}\right)\right) \cdot (t - t_0)}{1 + \left(\frac{\delta_1}{\delta_0} + \frac{\eta_1}{\eta_0}\right) \cdot (t - t_0)}} \tag{16}$$

This simple model can be applied to describe an early stage of sinusoidal growth of a perturbed thin shell, where the evolution of the TSI is not far from the initial conditions.